\documentclass[prd,aps,preprint,tightenlines,nofootinbib,superscriptaddress]{revtex4}

\usepackage{amsmath}
\usepackage{epsfig}

\newcommand{\insertfig}[2]{\mbox{\epsfxsize=#1cm \epsfbox{#2.eps}}}
\def\Journal#1#2#3#4{{#1} {#2} (#4) #3 }
\def\Journal#1#2#3#4{{#1} {#2} (#4) #3 }

\def\PRL{\em Phys. Rev. Lett.}

\begin{document}

\hfill{NPAC-08-24}

\title{\mbox{\hspace*{-0.35cm}\large Bino-driven electroweak baryogenesis with} \
 \mbox{\hspace*{-0.35cm}\large highly suppressed Electric Dipole Moments}}

\author{Yingchuan Li}
\email{yli@physics.umd.edu} \affiliation{Department of Physics,
University of Wisconsin, Madison, Wisconsin 53706 USA}
\author{Stefano Profumo }
\email{profumo@scipp.ucsc.edu} \affiliation{Department of Physics
and Santa Cruz Institute for Particle Physics, University of
California, 1156 High St., Santa Cruz, CA 95064, USA}
\author{Michael Ramsey-Musolf}
\email{mjrm@physics.wisc.edu} \affiliation{Department of Physics,
University of Wisconsin, Madison, Wisconsin 53706 USA}
\affiliation{Kellogg Radiation Laboratory, California Institute of
Technology, Pasadena, CA 91125 USA}
\date{\today}

\begin{abstract}

\noindent It is conventional wisdom that successful electroweak
baryogenesis in the Minimal Supersymmetric extension of the Standard Model (MSSM) is in tension with the non-observation of
electric dipole moments (EDMs), since the level of CP-violation responsible
for electroweak baryogenesis is believed to generate unavoidably
large EDMs. We show that CP-violation in the bino-Higgsino sector of the MSSM can account for successful
electroweak baryogenesis without inducing large
EDMs. This observation weakens the correlation between electroweak
baryogenesis and EDMs, and makes the bino-driven electroweak
baryogenesis scenario the least constrained by EDM limits.
Taking this observation together with the requirement of a
strongly first-order electroweak phase transition, we
argue that a bino-driven scenario with a light stop is the most phenomenologically viable MSSM electroweak baryogenesis scenario.

\end{abstract}

\maketitle

\section{introduction}

Explaining the origin of the observed baryon asymmetry of the
Universe (BAU) \cite{Komatsu:2008hk} is one of the most compelling problems
at the interface of cosmology, particle physics and nuclear physics. Among
baryogenesis scenarios, electroweak baryogenesis
(EWB)  \cite{ewb} is particularly subject to current and planned experimental scrutiny, given 
its essential dependence on new physics at the electroweak scale. It is well known that the standard model (SM)
cannot explain the observed BAU\cite{Farrar:1993hn}, even though it contains in principle all the necessary ingredients for successful
 baryogenesis \cite{Sakharov:1967dj}. In particular, the SM Higgs sector  does not
generate a first-order electroweak phase transition, while the CP-violating interactions of the SM would not generate
sufficiently large particle-antiparticle asymmetries at electroweak temperatures even if there was a strong first order SM phase transition. 
Therefore, successful EWB  requires 
new physics at the electroweak scale.


The Minimal Supersymmetric extension to the Standard Model (MSSM), a theoretical
framework that successfully addresses the naturalness problem of the SM, can also encompass a viable EWB mechanism for the generation of the BAU \cite{MSSMEWB,Carena:1997gx,Lee:2004we}. It
has been shown that the phase transition in MSSM can be strongly
first-order with a light, mainly right-handed, scalar top (stop)
\cite{lss,Carena:2008rt}. Moreover, the MSSM provides additional CP-violating
sources that may generate sufficiently large CP-violating asymmetries in the context of EWB. In general, however,  the non-observation
of permanent electric dipole moments (EDMs) places severe constraints on new
electroweak scale CP-violating interactions such as those of the MSSM. Specifically, the
current experimental bounds on the EDM of the electron, neutron, and the Mercury atom
($^{199}$Hg) are comparatively tight and constraining: $|d_e|<1.6\times 10^{-27} e
~{\rm cm}$ (90\% C.L.) \cite{Regan:ta}, $|d_n|< 2.9\times 10^{-26} e
~{\rm cm}$ (90\% C.L.) \cite{baker06}, and $|d_A(^{199}\textrm{Hg})|
< 2.1\times 10^{-28}e ~{\rm cm}$ (95\% C.L.) \cite{Romalis:2000mg}
(For recent reviews of EDM searches and their implications for MSSM,
see, {\em e.g.} Refs.~\cite{Pospelov:2005pr,RamseyMusolf:2006vr,Ellis:2008zy}). 
These results imply that complex CP-violating phases in the MSSM that generate one-loop EDMs must be
tiny compared to na\"ive expectations, leading to the so-called supersymmetric \lq\lq CP problem".  
The next generation of experiments on EDM searches will improve the current
sensitivity by two or more orders of magnitude \cite{EDMfuture}, and null results would only exacerbate the puzzle.

Solutions to the  supersymmetric CP problem, as well as to the supersymmetric flavor problem, have inspired
numerous theoretical studies and the formulation of specific frameworks where those issues are alleviated, such as ``more minimal'' SUSY
\cite{Cohen:1996vb} and \lq\lq split-SUSY" \cite{split}. For instance, in the latter
scenario one-loop EDM contributions are suppressed by the mass scale of the relevant scalar fermions. However, it has been realized that 
two-loop EDM contributions survive, and that they play, both in the split-SUSY scenarios and in others where sfermions are heavy, a dominant role in
constraining CP-violation in the MSSM \cite{Chang:2002ex,Pilaftsis:2002fe,Giudice:2005rz,Chang:2005ac,Li:2008kz}. 

On general grounds, one would expect that any large CP-violating source in the MSSM that is able to 
generate the BAU during the electroweak phase transition might also
induce large two-loop EDMs. In what follows, we show that there exists an important exception to this expectation, namely, CP-violating
interactions involving the relative phase between the supersymmetric Higgs-Higgsino mass term $\mu$ and the soft supersymmetry-breaking masses $M_1$ of the bino and $b$ of the Higgsino. We show that this phase, $\phi_1\equiv\mathrm{Arg}(\mu M_1 b^\ast)$, is essentially unconstrained by EDM measurements even at the two-loop level and that the associated CP-violating interactions may generate the observed BAU during the supersymmetric electroweak phase transition. On the other hand, the phase $\phi_2\equiv\mathrm{Arg}(\mu M_2 b^\ast)$, involving the wino supersymmetry-breaking mass $M_2$, induces large two-loop EDMs for sub-TeV superpartner masses and, thus, must be kept small in order to be consistent with experimental limits. Assuming SUSY is discovered at the Large Hadron Collider, successful EWB could still occur in the MSSM if it is driven by CP-violating bino-Higgsino interactions (rather than wino-Higgsino interactions) in the presence of a light nearly right-handed stop. This \lq\lq bino-driven"  (or \lq\lq neutralino-driven") EWB scenario (where $|M_1|\sim|\mu|$) requires a non-universality of the bino and wino phases relative to $\mu$ ($\phi_1\not=\phi_2$). We argue below that, while not generic, this situation occurs in well-motivated models of supersymmetry breaking.


A number of recent studies, including Ref.~\cite{Pilaftsis:2002fe,Chang:2002ex,Balazs:2004ae,Cirigliano:2006dg}, have addressed the interplay between EWB and EDMs in the MSSM. While some of them
\cite{Pilaftsis:2002fe,Chang:2002ex,Balazs:2004ae} concentrate on
the chargino-driven EWB scenario only, Ref.~\cite{Cirigliano:2006dg}, although dealing with both chargino-driven
and bino-driven EWB, assumed the same value for the bino and the
wino relative phases. To our knowledge, the scenario of bino-driven EWB with highly
suppressed EDMs  introduced here has not been discussed previously.
As a further motivation to investigate this framework, we recently completed and presented in  \cite{Li:2008kz} the complete calculation of the two-loop
chargino-neutralino contributions to EDMs, which play a vital role
in the interplay between EWB and EDMs. The results of this calculation  enable us to draw reliable conclusions on how the bino phase contributes to the EDMs, and
therefore to provide a solid test ground for the scenario of bino-driven EWB with
highly suppressed EDMs.

Our study is organized as follows: In section \ref{sec:scenario} we describe the specific pattern
of masses and phases that characterizes bino-driven EWB, and motivate why we expect highly suppressed EDMs,
followed by our numerical results. We then devote section III to our summary and conclusions.

\section{A scenario of successful EWB with highly suppressed EDMs}
\label{sec:scenario}

\subsection{ EWB requirements on MSSM parameters}

The requirement of a strongly first-order electroweak phase transition
is satisfied, in the context of the MSSM, in the light stop scenario \cite{lss},where the mass of the lighter, mostly right-handed, stop is less than 125 GeV, according
to the most recent analysis using renormalization group improved
effective potentials \cite{Carena:2008rt}. The masses of the
first-two generations of squarks and sleptons are kept heavier than
a few TeV to avoid the supersymmetric flavor and CP problems
\cite{Cohen:1996vb}. The heavier stop (mainly left-handed) also
needs to be heavier than a few TeV to satisfy the current Higgs mass
bound, and to suppress contributions to electroweak precision observables \cite{pdg}. In addition, the gluino mass should be larger than
about 500 GeV in order not to suppress the improvement on the first order character of the electroweak phase
transition \cite{Carena:2008rt}.

In contrast, Higgsinos, binos and winos must remain light to
trigger the needed CP-violating currents. Theoretical
studies show that, for specific mass patterns, the CP asymmetry in MSSM EWB can be resonantly
enhanced \cite{Carena:1997gx,Lee:2004we}. Unless the relevant particle masses are extremely light, the resonant enhancement of CP-violating sources is required to reproduce the observed BAU. This leads to two
scenarios. In the first one, the resonant enhancement occurs because the
Higgsino mass scale is close to either the bino or the wino soft
supersymmetry breaking masses,
$|\mu| \approx |M_1| $ or $ |\mu| \approx |M_2| $, corresponding to the
so-called
bino-driven and wino-driven EWB scenarios, respectively. In the second
one, instead, the resonant enhancement occurs because the soft 
supersymmetry breaking masses of
the right-handed and of the left-handed stops are close to each other, $
m^2_{\tilde{t}_R} \approx m^2_{\tilde{t}_L}$. The latter resonant condition
is, however, inconsistent with the simultaneous requirements of a light right-handed stop, as required by a strongly first-order phase transition, and
of a heavy left-handed stop, as needed by the current Higgs mass limit and precision electroweak data. Therefore, on general grounds we regard the first scenario only, either  involving bino- or wino-driven EWB, or even both, as being the phenomenologically viable and relevant one.

The CP-violating sources in the bino- or wino-driven EWB scenario live in the
chargino-neutralino sector. While numerous CP-violating phases appear in the most general MSSM parametrization, field redefinitions
can be employed to rotate away all but two physical phases in the chargino-neutralino sector. We take these phases to be the $\phi_{1,2}$ introduced above.
We will refer to
$\phi_1$ as the phase of the bino soft supersymmetry breaking mass, and to $\phi_2$ as the phase of the
wino mass, although they are indeed combinations of phases of the
Higgsino mass $\mu$, the gaugino mass $M_{1,2}$, and the soft Higgs mass term
$b$. A large enough phase $\phi_1$ or $\phi_2$ is needed, in
addition to the above-mentioned resonant condition on the masses, for successful EWB in the context of the bino-driven and
wino-driven scenarios, respectively. In addition, the baryon
asymmetry generated from MSSM EWB depends linearly on the relative
variation of the two Higgs fields along the bubble walls, $\Delta
\beta$, which receives significant suppression as the mass scale of
CP-odd Higgs, $m_A$, increases \cite{Moreno:1998bq}.

\subsection{Suppressed EDMs with viable MSSM EWB}

\begin{table}[!b]
\caption{Summary of the phases entering in CP-odd operators, and
of the conditions needed to suppress the operator without spoiling successful EWB. The symbol
$\phi_{\tilde{f}}$ denotes generic CP-violating phases in the
squark and slepton sector. $\phi_i \equiv {\rm arg}(\mu M_i
b^*)$ indicate the physical phases in the chargino-neutralino, and in the gluino sector.
Lastly, $m_{\tilde{f}_{1,2}}$ represents the (common) soft supersymmetry breaking masses of the first-two
generations of sfermions.}
\begin{center}
\label{tab:suppression}
\begin{tabular}{|c|c|c|}
\hline CP-odd operator & phases & suppression conditions without spoiling EWB\\
\hline $C^{4f}$ &  $\phi_1$, $\phi_2$, $\phi_3$, $\phi_{\tilde{f}}$
&
${\rm tan}\beta <$ 30 \\
\hline $d^{G}$ & $\phi_3$, $\phi_{\tilde{f}}$ &
${\rm sin}\phi_{\tilde{f}}< 10^{-2}$, ${\rm sin}\phi_3 < 10^{-2} $\\
\hline $d^{1-{\rm loop}}_{u,d,e}$ &  $\phi_1$, $\phi_2$, $\phi_3$,
$\phi_{\tilde{f}}$ &
$m_{\tilde{f}_{1,2}} >$ 10 TeV \\
\hline $\tilde{d}^{1-{\rm loop}}_{u,d}$ &  $\phi_1$, $\phi_2$,
$\phi_3$, $\phi_{\tilde{f}}$ &
$m_{\tilde{f}_{1,2}} >$ 10 TeV  \\
\hline $d^{2-{\rm loop}}_{u,d,e}(\tilde{t},\tilde{b},\tilde{\tau})$
& $\phi_{\tilde{f}}$ &
${\rm sin}\phi_{\tilde{f}} < 10^{-2}$ \\
\hline $\tilde{d}^{2-{\rm loop}}_{u,d}(\tilde{t},\tilde{b})$ &
$\phi_{\tilde{f}}$ &
${\rm sin}\phi_{\tilde{f}} < 10^{-2}$ \\
\hline $d^{2-{\rm loop}}_{u,d,e}(\chi^{\pm,0})$ & $\phi_{1}$,
$\phi_{2}$ &
${\rm sin}\phi_2 < 10^{-2}$, ${\rm sin}\phi_1 \sim {\cal O}(1)$ \\
\hline
\end{tabular}
\end{center}
\end{table}

\begin{figure}[t]
\begin{center}
\mbox{
\begin{picture}(0,120)(110,0)

\put(-60,-23){\insertfig{12}{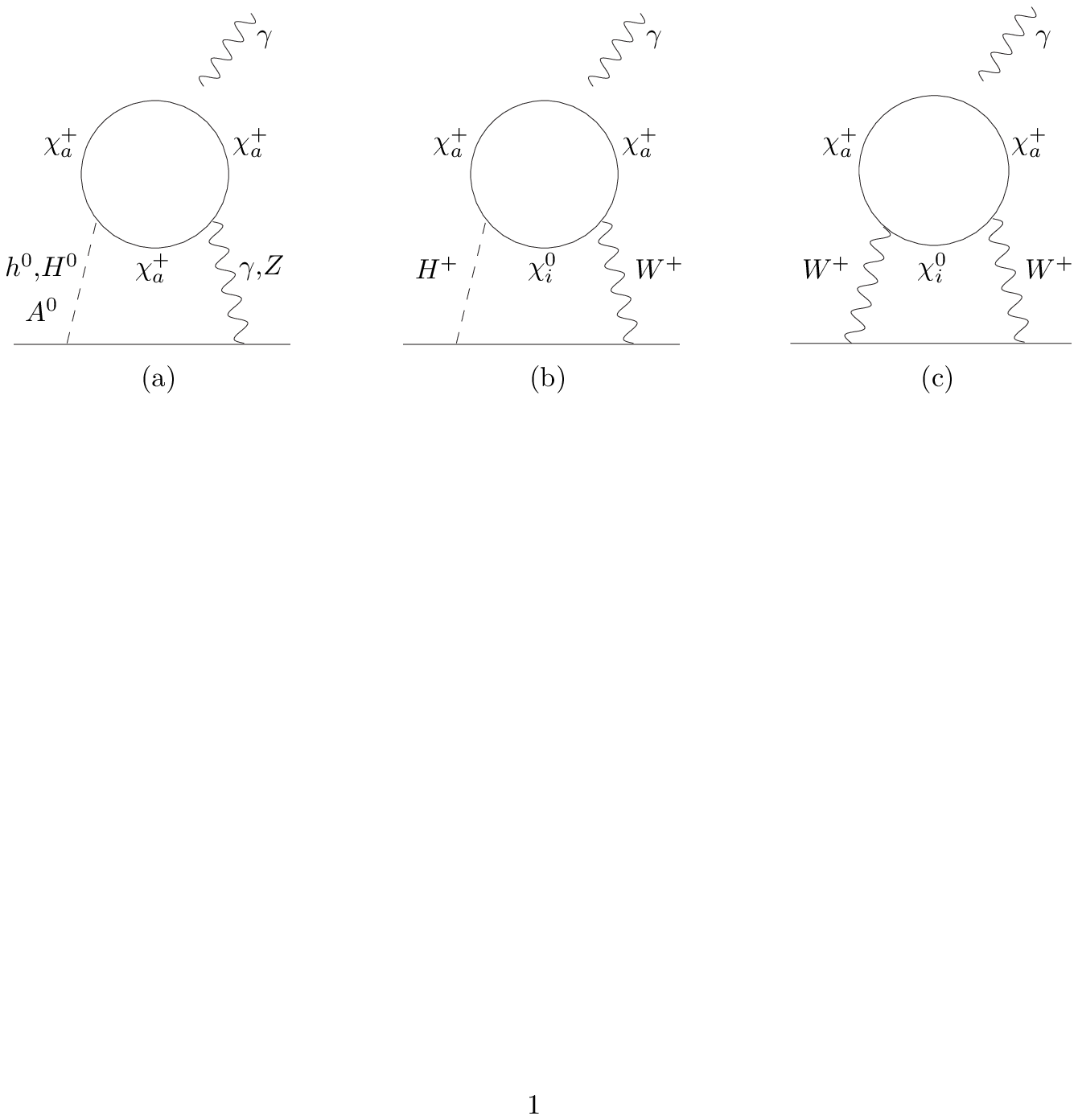}}

\end{picture}
}
\end{center}
\caption{The complete list of all chargino-neutralino two-loop diagrams contributing to EDMs of leptons and quarks. The external
photon line is attached to the charged particles in each diagram in all possible ways.
Mirror graphs are not displayed. \label{fig:LoopGraphs}}
\end{figure}

\begin{table}[!b]
\caption{Summary of mass scales and phases in the scenario
of successful bino-driven EWB with highly suppressed EDMs. The light
stop $\tilde{t}_1$ is predominantly right-handed, while the heavy
stop $\tilde{t}_2$ is mainly left-handed. The other mass scales and
phases are the same as in Table \ref{tab:suppression}.  The final entry gives the range for $\tan\beta$.}
\begin{center}
\label{tab:scenario}
\begin{tabular}{|c|c|c|c|c|c|c|c|c|c|}
\hline ${\rm sin}\phi_1$ &  ${\rm sin}\phi_{2,3}$,~${\rm
sin}\phi_{\tilde{f}}$ &
$|\mu| \approx |M_1|$, $|M_2|$, $m_A$ & $|M_3|$ & $m_{\tilde{f}_{1,2}}$, $m_{\tilde{t}_2}$ & $m_{\tilde{t}_1}$ & ${\rm tan} \beta$ \\
\hline $\sim {\cal O}(1)$ & $< {\cal O}(0.01)$ & $\sim$ few 100 GeV
& $>$ 500 GeV &
$>$ 10 TeV & $<$ 125 GeV & (3,30)  \\
\hline
\end{tabular}
\end{center}
\end{table}

The current most stringent EDM bounds are for the neutron,  the Thallium atom
($^{205}{\rm Tl}$), and the Mercury atom ($^{199}{\rm Hg}$). In general, they
receive contributions from operators associated with the lepton and quark EDMs, 
$d_{u,d,e}$; quark chromo-EDMs  $\tilde{d}_{u,d}$;
CP-odd 3-gluon Weinberg interaction, $d^{G}$ \cite{3Gop,Ellis:2008zy}; and CP-odd four-fermion
interactions, $C^{4f}$ (see e.g.~\cite{RamseyMusolf:2006vr} and \cite{Ellis:2008zy} for recent reviews). As shown in Ref.
\cite{Pospelov:2005pr,Ellis:2008zy}, the Thallium EDM is dominated
by the electron EDM operator $d_e$, and possibly by the four-fermion operator
$C^{4f}$ if ${\rm tan}\beta > 30$; the neutron EDM mainly stems from
the EDM and chromo-EDM operators of $u$ and $d$ quarks, $d_{u,d}$ and
$\tilde{d}_{u,d}$, and from the 3-gluon interaction $d^G$;  lastly, the Mercury EDM is
generated primarily by the chromo-EDM operators $\tilde{d}_{u,d}$.

These CP-violating operators are induced by various (physical) CP-violating phases in the MSSM, including  $\phi_{1,2}$ in the chargino-neutralino sector;
$\phi_3 \equiv \mathrm{Arg} (\mu M_3 b^*)$ in the gluino sector; and, lastly, in the sfermion-Higgs sector, $\mathrm{Arg}(\mu^* {\rm tan}\beta + A_f)$ and $\mathrm{Arg}(\mu^* {\rm
cot}\beta + A_f)$ for down- and up-type sfermions, respectively, which we generally refer to as $\phi_{\tilde{f}}$,
(where $y_f A_f$ is the coefficient of the supersymmetry-breaking triscalar interactions with $y_f$ being the fermion $f$ Yukawa coupling). We summarize in Table \ref{tab:suppression} the phases entering each CP-odd
operator. We also list the conditions under which  the corresponding CP-odd operator is suppressed
without affecting EWB.

The Higgs-mediated CP-odd 4-fermion operators $C^{4f}$ are only
enhanced at large ${\rm tan}\beta$ due to their ${\rm tan}^3\beta$
dependence \cite{Lebedev:2002ne}. By restricting to the ${\rm
tan}\beta < 30$ region (as also implied in the context of successful MSSM EWB by the recent study of Ref.~\cite{Carena:2008rt}), we keep this contribution small, and the
experimental bound on the Thallium EDM can be taken directly, in this regime, as a bound on $d_e$. (Incidentally, keeping ${\rm tan}\beta$ not too large also helps to
suppress other EDM contributions.) The CP-odd 3-gluon operator $d^G$
depends on the CP-violating phases in the sfermion sector,
$\phi_{\tilde{f}}$, and in the gluino sector, $\phi_3$, and it can be
suppressed by restricting these phases to be less than $10^{-2}$ \cite{Ellis:2008zy}. As discussed above, these phases are not crucial  to successful EWB.
With these operators suppressed, the remaining CP-odd operators are
the EDMs of leptons and quarks, as well as the chromo-EDMs of quarks.

The lowest order contributions to EDM and chromo-EDM operators are
induced at one-loop order \cite{Ibrahim:1997gj}. They involve the
first-two generations of sfermions, as well as neutralinos,
charginos and gluinos. Without affecting EWB, these one-loop
contributions are suppressed if the first-two generations of
sfermions are heavier than 10 TeV \cite{Cohen:1996vb,Lee:2004we,usinprep}. However, the
EDM constraints cannot be completely avoided by suppressing
one-loop contributions. It is well known that the two-loop
contributions of the Barr-Zee type \cite{Barr:1990vd} dominate over
one-loop contributions when the latter are suppressed by heavy sfermion
masses. Depending on the source of CP-violation, there
are two types of two-loop contributions. In one of them, the
CP-violation involves the third generation of sfermions
\cite{scalar}. Without affecting the EWB, these contributions to the two-loop
EDM and chromo-EDM contributions can be held below the experimental bounds
by suppressing the CP-violating phases in the sfermion sector, $\phi_{\tilde{f}}$,
as already employed to suppress the 3-gluon operator $d^G$. 

The second class of  two-loop contributions involves the CP-violating phases $\phi_{1,2}$ in the
chargino-neutralino sector. These are directly relevant to EWB,
since $\phi_{1,2}$-dependent interactions also generate CP asymmetries during the electroweak phase transition. These phases
contribute to the elementary fermion EDMs, but not to the chromo-EDMs. Moreover, both the CP-odd
Higgs and the charged Higgs, whose mass depends on the parameter $m_A$, enter the chargino-neutralino two-loop EDM contributions. This
provides yet another connection between this type of EDM contributions
and EWB: a crucial dependence on the same mass parameter $m_A$.

The complete set of chargino-neutralino two-loop diagrams that contribute to
quark and lepton EDMs in the MSSM are shown in Fig.~\ref{fig:LoopGraphs}. CP-violation stems from the chargino-neutralino loop, and is
propagated to quarks and leptons through the exchange of gauge and
Higgs boson pairs, including $\gamma h^0$, $\gamma H^0$, $Zh^0$,
$ZH^0$, $\gamma A^0$, $ZA^0$, and $WH^{\pm}$, or pure gauge boson
pairs which can only be $WW$. Notice that it cannot be transmitted through the
exchange of the neutral gauge boson pairs $\gamma \gamma$, $\gamma Z$, and $ZZ$
\cite{Giudice:2005rz}. Obviously, the bino phase $\phi_1$ can only
possibly enter the $WH^{\pm}$ and $WW$ contributions, since those are
the only ones that involve neutralinos.

A subset of the contributions to the chargino-neutralino two-loop EDMs have been studied in the past
\cite{Chang:2002ex,Pilaftsis:2002fe,Giudice:2005rz,Chang:2005ac},
and we recently presented the complete calculation in \cite{Li:2008kz},
making it possible to draw reliable conclusions on the correlation
between EWB and EDMs at the two-loop level. Without assuming gaugino mass unification,
we allow the phases $\phi_1$ and $\phi_2$ to be different. This is completely generic in the low-energy parametrization of softly broken supersymmetry in the MSSM, and in particular it occurs in some supersymmetry breaking
models such as ``mirage mediation'' \cite{mirage}, wherein gaugino
masses originate from more than one mediation mechanism, or ``gaugino mediation'' (see Ref.~\cite{Baer:2002by} and references therein). 

The main result of the present analysis is that the EDM contribution induced by $\phi_1$ is suppressed
compared to $\phi_2$ by a factor of $\sim$ 0.02. This suppression is due to
several effects:
\begin{enumerate}
\item While the wino phase $\phi_2$ enters all two-loop
contributions, the bino phase $\phi_1$ can only possibly enter the
$WH^{\pm}$ and $WW$ contributions, which, as shown in Ref.
\cite{Li:2008kz}, amounts to about $20\%$ of the total 2-loop EDM chargino-neutralino
contribution.
\item Wherever the bino enters in $WH^{\pm}$ and
$WW$ diagrams, its contribution is suppressed by a factor of $(g'/g)^2={\rm
tan}^2\theta_{W} \sim 0.3$ compared to the corresponding wino
contribution.
\item While the product of these two factors gives a
suppression factor of $0.06$, the further factor of $0.3$ needed to
explain the numerical result presumably stems from the fact that, in the
$WH^{\pm}$ and $WW$ contributions, the $W$ boson directly couples to
the wino, but not to the bino, and the latter only enters through its coupling to the
Higgs in $WH^{\pm}$, or through its mixing with Higgsino. 
\end{enumerate}
Since interactions involving
$\phi_1$ alone can generate enough baryon asymmetry in the
bino-driven EWB scenario, the weak dependence of EDMs on $\phi_1$
indicates the existence of a scenario for successful MSSM EWB consistent with highly
suppressed EDMs: the bino-driven scenario with a light, mainly right-handed
stop. This scenario is characterized by the specific pattern of MSSM masses
and phases summarized in Tab.~\ref{tab:scenario}.

\begin{figure}
\centerline{\epsfig{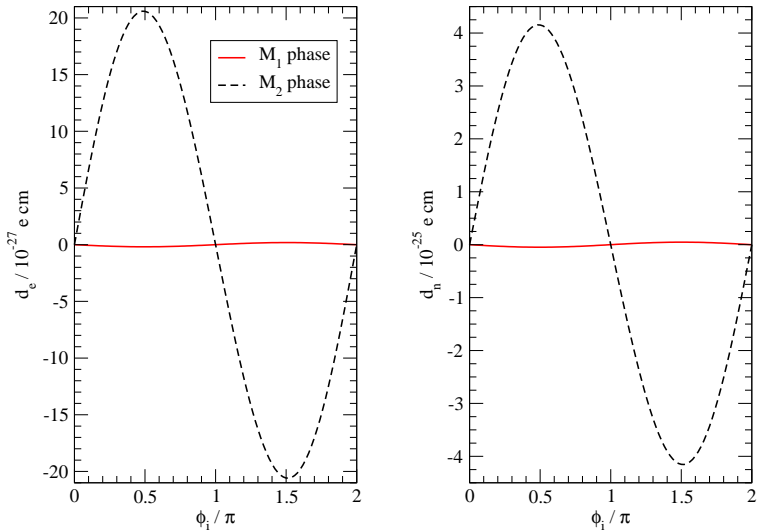}} \caption{The
electron (left) and neutron (right) electric dipole moment as a
function of the bino and wino phase.} \label{fig:phim}
\end{figure}

In order to show concrete numerical results for our scenario we choose, for
definiteness,  the following reference benchmark setup:
\begin{equation}
\mu=200\ {\rm GeV},~~|M_1|=95\ {\rm GeV},~~ |M_2|=190\ {\rm
GeV},~~\tan\beta=10,~~m_A=300\ {\rm GeV}. \label{eq:reference}
\end{equation}
This setup is consistent with (among other constraints) the limits from $b\to s\gamma$ \cite{bsg}. In Fig.
\ref{fig:phim}, we show the effect of a non-vanishing bino phase $\phi_1$ (red
lines) and wino phase $\phi_2$ (black dashed) on the electron (left)
and neutron (right) EDMs\footnote{The Mercury EDM is suppressed
in the parameter space region of interest here, as it is generated primarily by
the chromo-EDM operators, whose contributions from both one-loop and
two-loop are suppressed.}. The figure indicates clearly that the
size of the EDM contribution induced by $\phi_1$ is suppressed by a factor of $0.01 - 
0.02$ compared to that associated with $\phi_2$ .

\begin{figure}
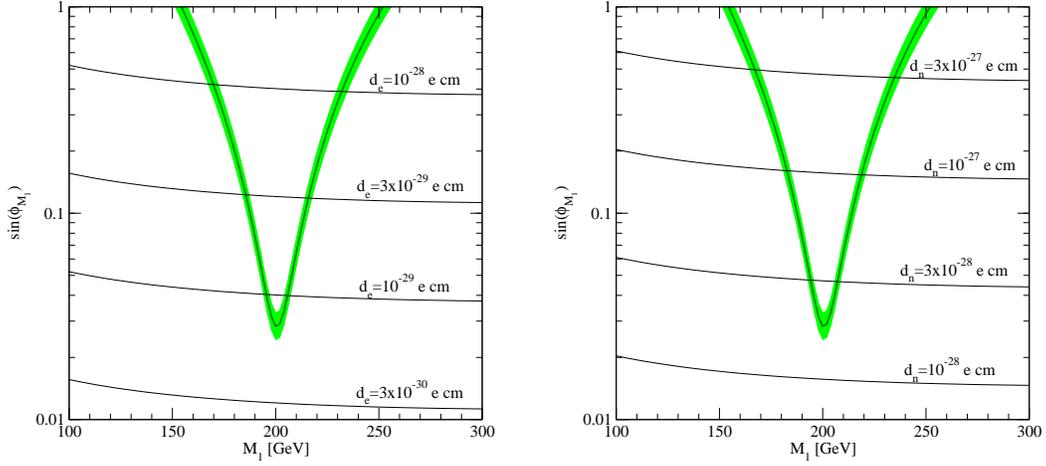

\centerline{\mbox{\epsfig{file=M1_de.eps,width=6.5cm,angle=0}\qquad\epsfig{file=M1_dn.eps,width=6.5cm,angle=0}}}
\caption{The green band shows the region, in the ($M_1,{\rm
sin}\phi_1$) plane compatible with electroweak baryogenesis. We
assume that ${\rm sin}\phi_2=0$. On the same plane, we indicate
iso-level curves at constant values for the electron (left) and for
the neutron (right) EDMs.} \label{fig:m1}
\end{figure}

\begin{figure}
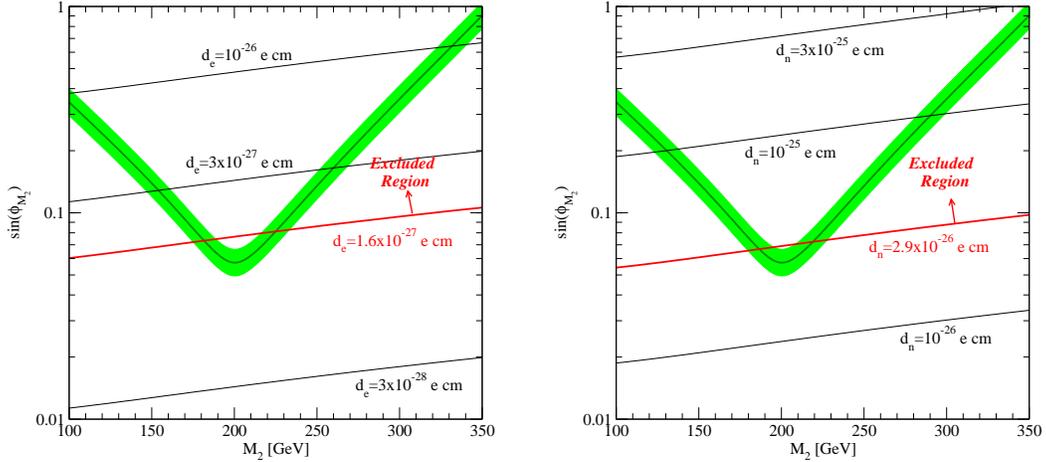

\centerline{\mbox{\epsfig{file=M2_de.eps,width=6.5cm,angle=0}\qquad\epsfig{file=M2_dn.eps,width=6.5cm,angle=0}}}
\caption{The green band shows the region, in the ($M_2,{\rm
sin}\phi_2$) plane compatible with electroweak baryogenesis. We
assume that ${\rm sin}\phi_1=0$. On the same plane, we indicate
iso-level curves at constant values for the electron (left) and for
the neutron (right) EDMs. Parameter space points above the red lines
are excluded by current experimental constraints on electron and
neutron EDMs.} \label{fig:m2}
\end{figure}

The significantly different impact on the size of the induced EDMs for $\phi_1$ versus
$\phi_2$ makes the bino-driven EWB scenario much less constrained by EDM
bounds than the wino-driven option. This is illustrated in detail in Fig.
\ref{fig:m1} and Fig. \ref{fig:m2}, where we compare bino-driven and wino-driven EWB by showing the predicted BAU, as well as curves of constant
 electron and neutron EDMs on the ($|M_1|$,$\phi_1$) and
($|M_2|$,$\phi_2$) planes, respectively. Again, for definiteness, we
keep $|M_2|=2|M_1|$ and set the other parameters to the values indicated in Eq.~(\ref{eq:reference}). 

In Fig. \ref{fig:m1} and Fig. \ref{fig:m2}, the green bands indicate the
region compatible with the production of a baryon asymmetry ${\rm
Y_B}=9.2\times 10^{-11}$ at the 5-$\sigma$ level (according to the
results reported in Ref.~\cite{Komatsu:2008hk}). We observe that as
$|M_1|$($|M_2|$) approaches $|\mu|=200$ GeV,  the resonant enhancement
becomes larger and larger, and thus the phase $\phi_1$($\phi_2$) needed to
generate enough baryon asymmetry becomes smaller and smaller (no enhancement occurs in the two-loop EDMs if $|\mu|\sim|M_{1,2}|$). In turn, this makes it easier to
evade the EDM bounds. However, for the reasons outlined above, one sees that, since the $\phi_1$
contribution to EDMs is much smaller than that from $\phi_2$, all
the values of ${\rm sin}\phi_1$ are presently consistent with experimental EDM bounds, while the range of viable ${\rm
sin}\phi_2$ values is constrained to a very limited parameter space (and likely ruled out when a more realistic Higgs profile is used). Future neutron and electron EDM searches with $\sim 100$ times better
sensitivity than existing experiments would be needed to fully explore the
CP-violating parameter space in the presently proposed bino-driven EWB
scenario.

We note that the BAU-allowed bands have been obtained from the work of Ref.~\cite{Lee:2004we}, which included the effects
of both resonantly-enhanced chiral relaxation and CP-violating sources in the bino-driven and wino-driven regimes for a simple, step-function wall profile. Had we employed a more realistic profile, leading to a somewhat smaller BAU (see, {\em e.g.},  Ref.~\cite{Chung:2008aya}), the BAU-compatible regions in Figs. \ref{fig:m1} and \ref{fig:m2} would have moved to even larger values of the CP-violating phases corresponding to larger predicted magnitudes for the EDMs. In this respect,  Figs. \ref{fig:m1} and \ref{fig:m2} give the most optimistic expectations for the wino-driven scenario, whose viability is clearly marginal. In contrast,  the bino-driven scenario would  be still be easily compatible with the observed BAU and present EDM limits when a more realistic profile is employed and the full set of transport equations are solved numerically, as in Ref.~ \cite{Chung:2008aya}. Consequently, we rely here on the simpler, schematic solution as it adequately addresses our primary point. 

\section{summary and conclusions}

We have presented a novel possibility for reconciling present and prospective experimental limits on the EDMs of elementary particles with successful EWB in the MSSM. We pointed out that the most relevant CP violating phases for EWB
are the bino phase $\phi_1$ and the wino phase $\phi_2$. We showed that, with
its impact on EDMs suppressed by about two orders of magnitude compared to
that of the wino phase $\phi_2$, the bino phase $\phi_1$ is only weakly
constrained by the EDM bounds, and can be of order one. Since the bino
phase by itself can generate the observed BAU in the bino-driven EWB
scenario, our analysis revealed that bino-driven EWB is a scenario with
the least tension with EDM constraints. This conclusion is unambiguously
supported by the numerical results we presented. We leave the detailed study of the interplay between EWB and EDM over a larger cross section of the MSSM parameter space to a more comprehensive future study
\cite{usinprep}.

Besides the CP violation requirement, the other element needed in the MSSM for successful
EWB is a strongly first-order phase transition, which leads to the additional requirement of a light stop \cite{lss,Carena:2008rt}. We therefore argue that bino-driven EWB with a non-universal gaugino-Higgsino CP-violating phase and with a light stop is the most promising scenario for successful EWB in the MSSM. Interestingly, we notice as a last comment that the specific mass spectrum and CP violating phases needed in this scenario will also be tested with colliders and explored in dark matter searches in the near future \cite{Cirigliano:2006dg}.

\section*{Acknowledgements}
This work was supported U.S. Department of Energy Contracts DE-FG02-08ER41531(MRM and YL)
and DE-FG02-04ER41268 (SP) and by the Wisconsin Alumni Research Foundation (MRM and YL). MRM thanks the Department of Energy's Institute for Nuclear Theory at
the University of Washington for its hospitality  during the completion
of this work.

\end{document}